\documentclass[traditabstract]{aa}

\usepackage{graphicx}
\usepackage{txfonts}
\usepackage{natbib}
\bibpunct{(}{)}{;}{a}{}{,} 

\begin{document}

\title{Is Gliese 581d habitable? Some constraints from radiative-convective climate modeling}

   \author{R. D. Wordsworth
          \inst{1}
\fnmsep\thanks{rwlmd@lmd.jussieu.fr}
          \and
          F. Forget
	\inst{1}
	\and
          F. Selsis
	\inst{2,3}
	\and
          J.-B. Madeleine
	\inst{1}
	\and
          E. Millour
	\inst{1}
	\and
          V. Eymet
	\inst{4}
          }

   \institute{Laboratoire de M\'et\'erologie Dynamique, Institut Pierre Simon Laplace, Paris, France
         \and
            Universit\'e de Bordeaux, Observatoire Aquitain des Sciences de  l'’Univers,
2 rue de l'’Observatoire, BP 89, F-33271 Floirac Cedex, France
\and
CNRS, UMR 5804, Laboratoire d’'Astrophysique de Bordeaux,
2 rue de l’'Observatoire, BP 89, F-33271 Floirac Cedex, France
	\and
Laboratoire Plasma et Conversion de l'Energie, Universit\'e Paul Sabatier, Toulouse, France
             }

   \date{Received May 27, 2010}

  \abstract
   {The recently discovered exoplanet Gl581d is extremely close to the outer edge of its system's habitable zone, which has led to much speculation on its possible climate. We have performed a range of simulations  to assess whether, given  simple  combinations of chemically stable greenhouse gases, the planet could sustain liquid water on its surface. For best estimates of the surface gravity, surface albedo and cloud coverage, we find that less than 10 bars of CO$_2$ is sufficient to maintain a global mean temperature above the melting point of water. Furthermore, even with the most conservative choices of these parameters, we calculate  temperatures  above the water melting point  for CO$_2$ partial pressures greater than about 40 bar. However, we note that as Gl581d is probably in a tidally resonant orbit, further simulations in 3D are required to test whether such atmospheric conditions are stable against the collapse of CO$_2$ on the surface.} 


\titlerunning{Constraints on the climate of Gliese 581d}
   \maketitle

\section{Introduction}

In 2007, radial velocity measurements were used to discover two new planets in the Gl581 system \citep{udry2007}. These planets have captured much attention both in the community and among the general public, as their minimum masses were measured to be below 10 $M_{Earth}$, and they are close to the edges of their system's nominal `habitable zone', i.e., the loosely defined orbital region in which planets can sustain liquid water on their surfaces. The first planet, Gl581c, which is closer to its star and was the first discovered, was initially estimated to be potentially habitable based on its equilibrium temperature $T_{eq}=320$ K, using an Earth-like planetary albedo 0.29.  In contrast, the second planet Gl581d  has an equilibrium temperature $T_{eq}= 195$ K for  an albedo of 0.2, which suggests it may be too cold to sustain surface liquid water. However, these analyses neglect any possible warming of the surface due to the planet's atmosphere.

In the first detailed assessment of the potential habitability of these planets, \citet{selsis2007} reviewed a variety of factors that could influence their climates. They concluded that based on standard assumptions of atmospheric warming by a mixture of CO$_2$ and H$_2$O (with possible regulation of CO$_2$ via the carbonate-silicate cycle), Gl581c was unlikely to be habitable, while for Gl581d,  the situation was much less clear. According to \citet{kasting1993}, the outer edge of the habitability zone is most likely the distance at which CO$_2$ condensation begins to occur on the surface of the planet. However, CO$_2$ condensation in the atmosphere leads to the formation of CO$_2$ clouds, which can cause a strong warming effect due to the scattering of infrared radiation \citep{forget1997}. Hence it was concluded that further climate simulations were required.

To investigate the possible climate of Gl581d under a range of conditions, we have performed one-dimensional radiative-convective  calculations. In Section \ref{sec:method} we discuss the model we used, while in Section \ref{sec:results} we present our results for varying atmospheric compositions, surface albedos, gravity and cloud coverage. We also present some simple three-dimensional simulations that highlight the limitations of the one-dimensional globally averaged approach. In Section \ref{sec:discuss}, we discuss the implications of our results and suggest directions for future research.

\section{Method}\label{sec:method}

Our radiative scheme is based on the correlated-$k$ model, with the absorption data calculated directly from high resolution spectra. We used the program \emph{kspectrum}\footnote{http://code.google.com/p/kspectrum/} to produce line-by-line spectra from the HITRAN 2008 database \citep{hitran2009}. These were then converted to correlated-\emph{k} coefficients for use in the radiative transfer calculations. The CO$_2$ collision-induced absorption (CIA) was parameterized using the results of \citet{baranov2004} and \citet{Gruszka1998}, while the sub-lorentzian profiles of \citet{Perrin1989} were used for the CO$_2$ far line absorption. For further details and justification of the method used, refer to \citet{Wordsworth2010}.  
Note that at  high CO$_2$ partial pressures (tens of bars), additional sources of CIA may appear that we have not been able to take into account  \citep{Eymet2009}. However, these would tend to increase the surface temperature in our calculations. As we are interested in making a conservative estimate of habitability, we can safely neglect them here.

A matrix of coefficients was produced on a 6 $\times$ 9 temperature and log-pressure grid $T = \{100, 150,\ldots, 350 \}$ K, $p  = \{ 10^{-3}, 10^{-2}, \ldots, 10^5 \} $ mbar. For mixed CO$_2$-H$_2$O atmospheres, we produced correlated-$k$ coefficients for H$_2$O volume mixing ratios $q_{H_2O}=\{10^{-7}, 10^{-6}, \ldots, 10^{-1} \}$. We used 32 spectral bands in the longwave and 36 in the shortwave. Sixteen points were used for the $g$-space integration, where $g$ is the cumulated distribution function of the absorption data for each band. For the stellar spectra, we used the Virtual Planet Laboratory AD Leo and solar data \citep{Segura2005} for M-class and G-class stars, respectively. 
The solar zenith angle $\theta_z$ was set to $60^\circ$, and the stellar flux was further divided by two to give a spatially and temporally averaged value $F_m=S_0/ 4\sqrt{1-e^2}$. 
As planetary albedo is a nonlinear function of zenith angle, the values
obtained in 1D can differ slightly from those produced by a
three-dimensional model. However, a $60^\circ$ zenith angle generally
overestimates the globally averaged value, and hence predicts lower
surface temperatures. We therefore neglected the error due to this
difference here.
Here $S_0 =381.4  $ W~m$^{-2}$  is the unaveraged stellar flux at the Gl581d semi-major axis $a$, and $e$ is the eccentricity. $F_m$ is given along with $a$, $e$ and the stellar luminosity $L$  in Table 1. The factor involving $e$ in the definition of $F_m$ comes from the fact that the  flux averaged over one orbit is higher in an eccentric orbit than in an equivalent circular one.  

To account for the radiative effects of both clouds and Rayleigh scattering in our simulations, we used the \citet{toon1989} scheme. Rayleigh scattering was included by the method described in \citet{Hansen1974}, while the properties of the clouds were computed from Mie theory. Refractive indices for the Mie calculation were taken from \citet{Hansen2005} and \citet{Warren1984} for the CO$_2$ and H$_2$O clouds, respectively. In the simulations with CO$_2$ clouds, a simple microphysical scheme was used to calculate the cloud opacity in each layer. Following \citet{Forget2004AGU}, the number of condensable nuclei per kilogram of air $N_{c}$ was treated as a free parameter. At each timestep, the mean cloud particle radius at each level was calculated from $N_{c}$ and the amount of condensed CO$_2$. This was then used to calculate the mean particle sedimentation rates and cloud radiative properties.

For water clouds, we used a simple approach. Fixed layers of particles of radius 10 $\mu$m and varying optical depth (defined at a reference wavelength 1.5 $\mu$m) were placed in the low troposphere, which we defined arbitrarily as the level at which $p=0.5 p_s$. In simulations with vertically varying amounts of water vapour, we used profiles of the form
\begin{equation}
q_{H2O}=q_{sat} RH
\end{equation}
where $q_{H2O}$ is the water vapour mixing ratio, $q_{sat}$ the water vapour saturation ratio, and $RH$ the relative humidity. The latter was defined as in \citet{Manabe1967}
\begin{equation}
RH = 0.77(p/p_{surf} - 0.02)/0.98\label{eq:RH}
\end{equation}
until $p/p_{surf}<0.02$, after which we set $q_{H2O}=0$.

To model convection in the lower atmosphere we used adjustment to keep the lapse rate $\Gamma$ stable. In dry ideal atmospheres, $\Gamma = -g \slash c_p$. As dense CO$_2$ is a non-ideal gas, $\Gamma$ can deviate from this value in the lower troposphere when the pressure is sufficiently high. We assess the importance of this deviation in the Appendix.  The effects of moist convection on the lapse rate were not included. While moist lapse rates are shallower than dry ones, which results in lower surface temperatures, the quantity of unsaturated water vapour in cold atmospheres is low. The error due to this approximation was  small in our calculations for surface temperatures below 273 K, and hence unimportant to our assessment of the minimum CO$_2$ partial pressure necessary for habitability. In contrast, the inclusion of CO$_2$ condensation above the troposphere was extremely important. We took this effect into account using the algorithm described in \citet{Forget1998} and  vapour pressure curves derived from \citet{CRC2000}.

Thirty vertical levels in standard sigma coordinates were used, and all simulations were iterated in time until a steady state had been reached (this took up to 50 Earth years of simulation time for the highest pressure runs). In all cases the energy balance (incoming vs. outgoing radiation) of the model was checked at the end of each run.  
A list of all parameters used in the simulations is given in Table~\ref{tab:params}.

\begin{table}[t]
\centering
\caption{Planetary and stellar properties used in the one-dimensional simulations. When multiple values are shown, those in bold are the standard values used.}
\begin{tabular}{lll}
\hline \hline
Stellar luminosity & $L$ [$L_{Sun}$] & 0.0135\\
Stellar spectrum &   & Sol, {\bf AD Leo}\\
Semi-major axis & $a$ [AU] & 0.22 \\
Eccentricity & $e$ & 0.38 \\
Averaged stellar flux & $F_{m}$ [W m$^{-2}$] & 103.1\\
Relative humidity & $RH$ & {\bf 0}, variable, 1.0 \\
Surface gravity & $g$ [m s$^{-1}$] & 10.0, {\bf 20.0}, 30.0 \\
Surface albedo & $A_s$ & 0.1, {\bf 0.2}, 0.3 \\
CO$_2$ partial pressure & $p_{CO_2}$ [bars] & {\bf 0.2 - 50} \\
N$_2$ partial pressure & $p_{N_2}$ [bars] & {\bf 0}, 1, 5, 20 \\
CO$_2$ cloud cover & $\tau_1 $ & {\bf none}, dynamic \\
H$_2$O cloud cover & $\tau_2 $ & {\bf none}, 1, 2, 3 \\
\hline \hline
\end{tabular}\label{tab:params}
\end{table}

\section{Results}\label{sec:results}

We have studied the effects of a range of climate parameters on the mean surface temperature of Gl581d.  For clarity,  this section is subdivided according to the effect studied. When not explicitly stated otherwise in the text, the parameters used for each simulation are those in bold in Table~1.

\subsection{Stellar spectrum}\label{subsec:stellar}

\begin{figure}[h]
	\begin{center}
		{\includegraphics[width=3.5in]{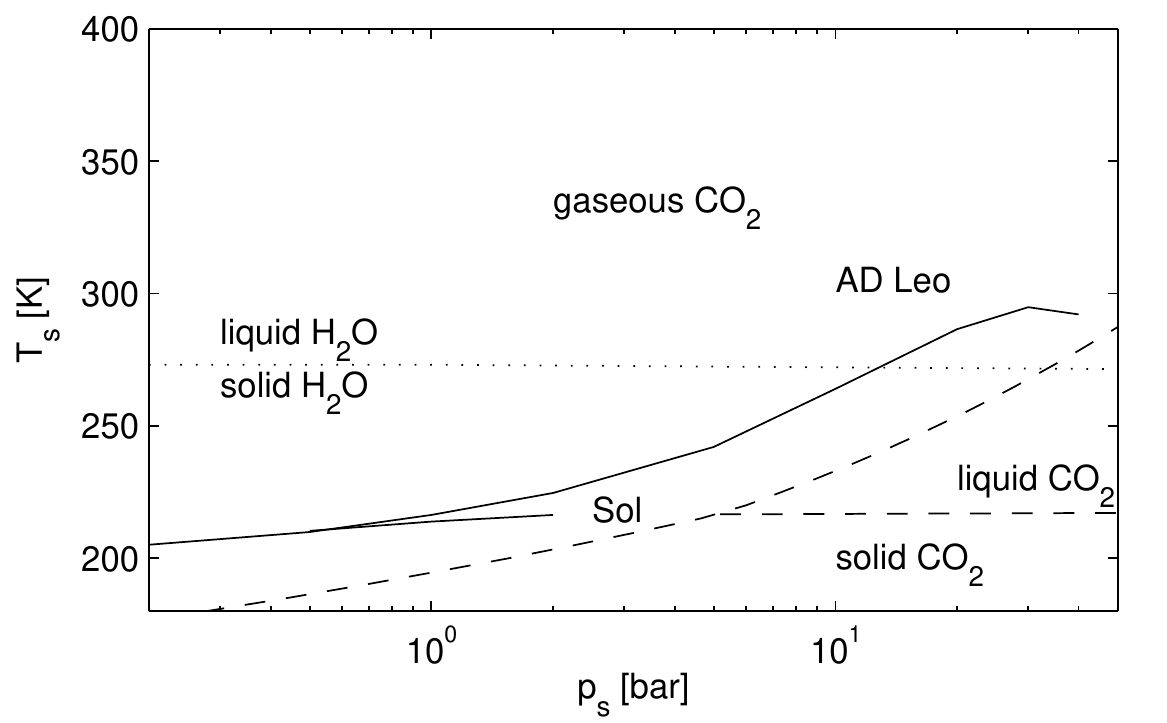}}
	\end{center}
	\caption{Surface temperature vs. surface pressure for clear pure CO$_2$ atmospheres under G-class (Sol) and M-class (AD Leo) stellar insolation. Dashed and dotted lines show CO$_2$ and H$_2$O phase transitions, respectively. In the G-class case, the line is discontinued because the atmosphere condenses on the surface for pressures greater than a few bar.}
	\label{fig:sspectra}
\end{figure}

To better understand the differences in climate caused by the fact that Gliese 581 is an M-class star, we first performed simulations comparing G-class (Sol) and M-class (AD Leo) stellar spectra. In both cases we normalized the total fluxes to the same value $F_m$ (see Table \ref{tab:params}). Figure \ref{fig:sspectra} shows the equilibrium mean surface temperatures obtained as a function of surface pressure for these two cases.

As can be seen, the clear pure CO$_2$ atmospheres under a G-class star collapse on the surface for pressures greater than about 3 bar, but when the star is M-class, temperatures continue to increase, reaching the water melting point at just over 10 bar and a maximum value at around 30 bar. The essential reason for this difference is that Rayleigh scattering, which has an optical depth $\tau_R \propto \lambda^{-4}$, has a much weaker effect on the red-shifted M-class stellar spectrum. 

\begin{figure}[h]
	\begin{center}
		{\includegraphics[width=3.5in]{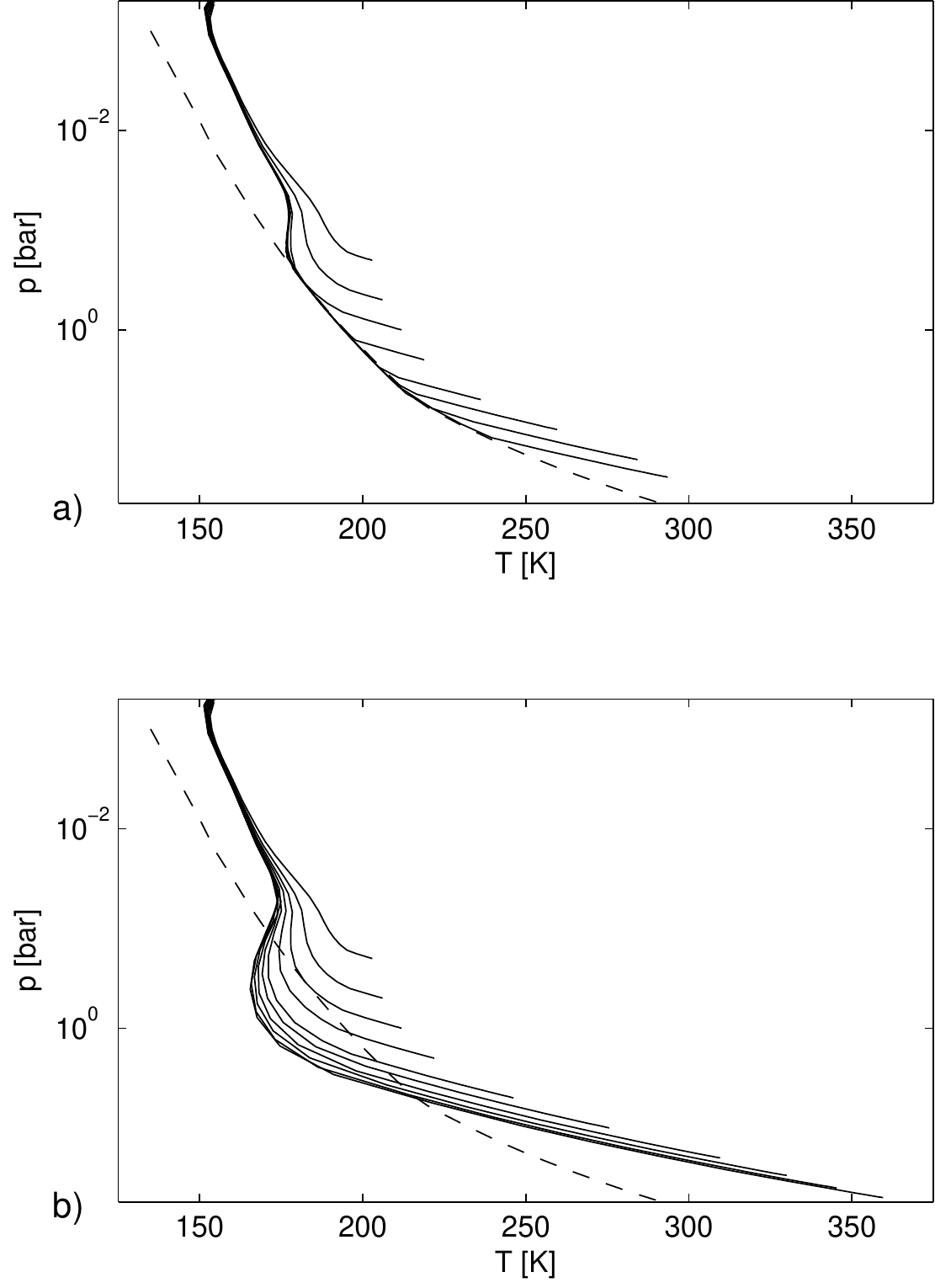}}
	\end{center}
	\caption{Steady-state vertical temperature profiles for clear pure CO$_2$ atmospheres of different surface pressures under M-class stellar insolation a) with and b) without CO$_2$ condensation included. The dashed line shows the CO$_2$ saturation pressure.}
	\label{fig:profiles}
\end{figure}

In addition to reducing the effect of Rayleigh scattering, the red shift in the stellar spectrum leads to increased warming in the upper stratosphere. This effect is clear from the vertical temperature profiles, which are plotted in Figure \ref{fig:profiles}. As can be seen, near-IR CO$_2$ absorption leads to a  temperature inversion near 0.05 bar, which becomes even more pronounced if CO$_2$ condensation is neglected in the calculation (Figure \ref{fig:profiles}b).

\begin{figure}[h]
	\begin{center}
		{\includegraphics[width=3.5in]{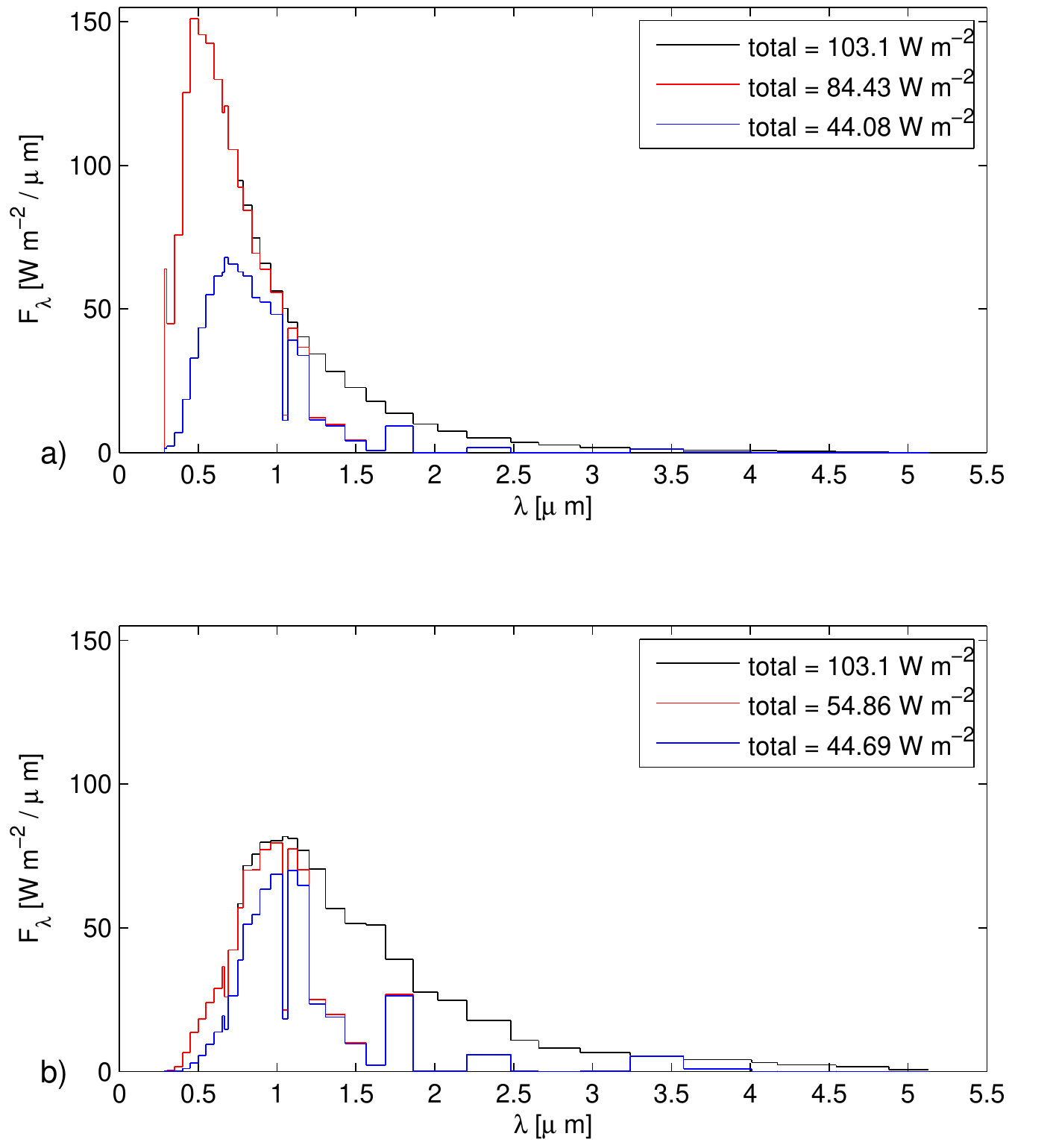}}
	\end{center}
	\caption{Incident flux in the visible at the top of the atmosphere (black) and at the ground, with (blue) and without (red) Rayleigh scattering included. In this example the atmosphere consisted of 40 bars  pure CO$_2$ (no clouds), while the stellar spectrum was G-class (Sol) for a) and M-class (AD Leo) for b).}
	\label{fig:2specs}
\end{figure}

To show the importance of these separate effects more clearly, in Figure \ref{fig:2specs} we have plotted the incident stellar spectra and the radiation reaching the ground for cases with and without Rayleigh scattering included. As can be seen, Rayleigh scattering dominates for wavelengths $<$ 1 $\mu$m, while gaseous absorption dominates above 1 $\mu$m.

As AD Leo has a mean temperature of around 3400 K, the planetary albedo values we obtain are lower than those for the blackbody 3700 K case that was used to define the habitable boundary in \citet{kasting1993}. This leads to a significant increase in the calculated surface temperature. Note that the temperature of Gl581 is approximately 3200 K, so the planetary albedos we calculate using the AD Leo spectrum should lead to conservative habitability estimates.

\subsection{CO$_2$ clouds}\label{subsec:clouds}

\begin{figure}[h]
	\begin{center}
		{\includegraphics[width=3.5in]{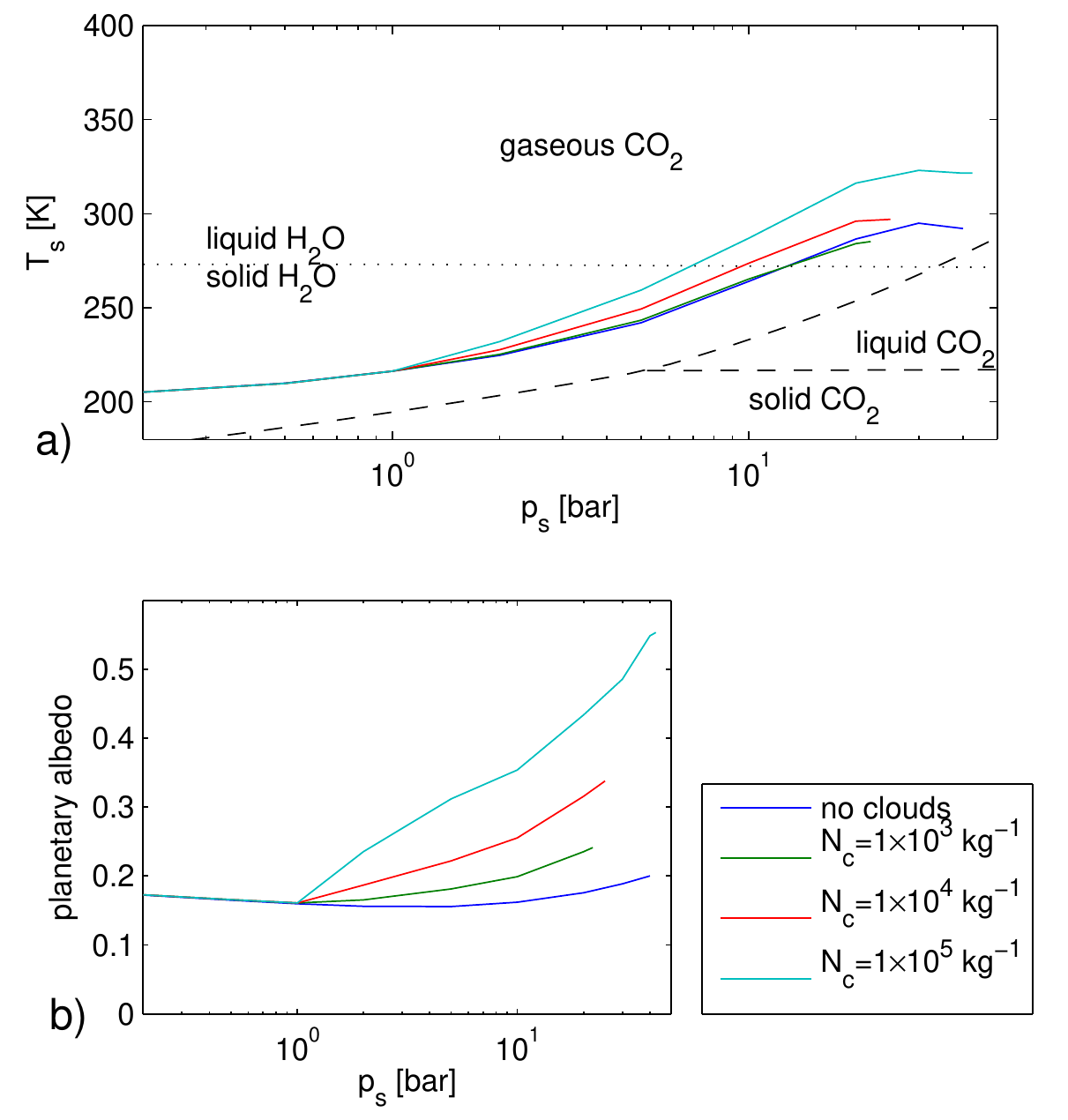}}
	\end{center}
	\caption{a) Surface temperature vs. surface pressure for pure CO$_2$ atmospheres under M-class insolation with CO$_2$ cloud radiative effects included. Dashed and dotted lines show CO$_2$ and H$_2$O phase transitions, respectively. b) Planetary albedo vs. surface pressure for the same simulations as in a).}
	\label{fig:co2clouds}
\end{figure}

If CO$_2$ condensation occurs in the upper atmosphere of a planet, it should cause CO$_2$ cloud formation. This effect is observed for example in the present-day Martian mesosphere \citep{Montmessin2007}. While there are still many unknowns associated with the microphysics of CO$_2$ clouds, a basic estimate of their effect can be gained through the scheme described in Section \ref{sec:method}. In Figure \ref{fig:co2clouds}a, the surface temperature vs. surface pressure is plotted for simulations with radiatively active CO$_2$ clouds and varying values of the condensation parameter $N_c$.   The infrared scattering effect described by \citet{forget1997} is expected to be slightly less efficient when the star is M-class, because a greater portion of the incident stellar radiation is also scattered back to space. Nonetheless, the clouds still increase the surface warming by an amount that increases with $N_c$, up to a theoretical maximum of  around 30 K at 40 bar for $N_c = 10^5$ kg$^{-1}$. Figure \ref{fig:co2clouds}b shows that at high pressures, the planetary albedo begins to increase significantly due to CO$_2$ cloud coverage. This ultimately limits the maximum warming effect that the clouds can provide, as the downward scattering of thermal radiation from the planet's surface becomes less important than the direct reflection of stellar radiation to space.

\subsection{Surface gravity}\label{subsec:gravity}

\begin{figure}[h]
	\begin{center}
		{\includegraphics[width=3.5in]{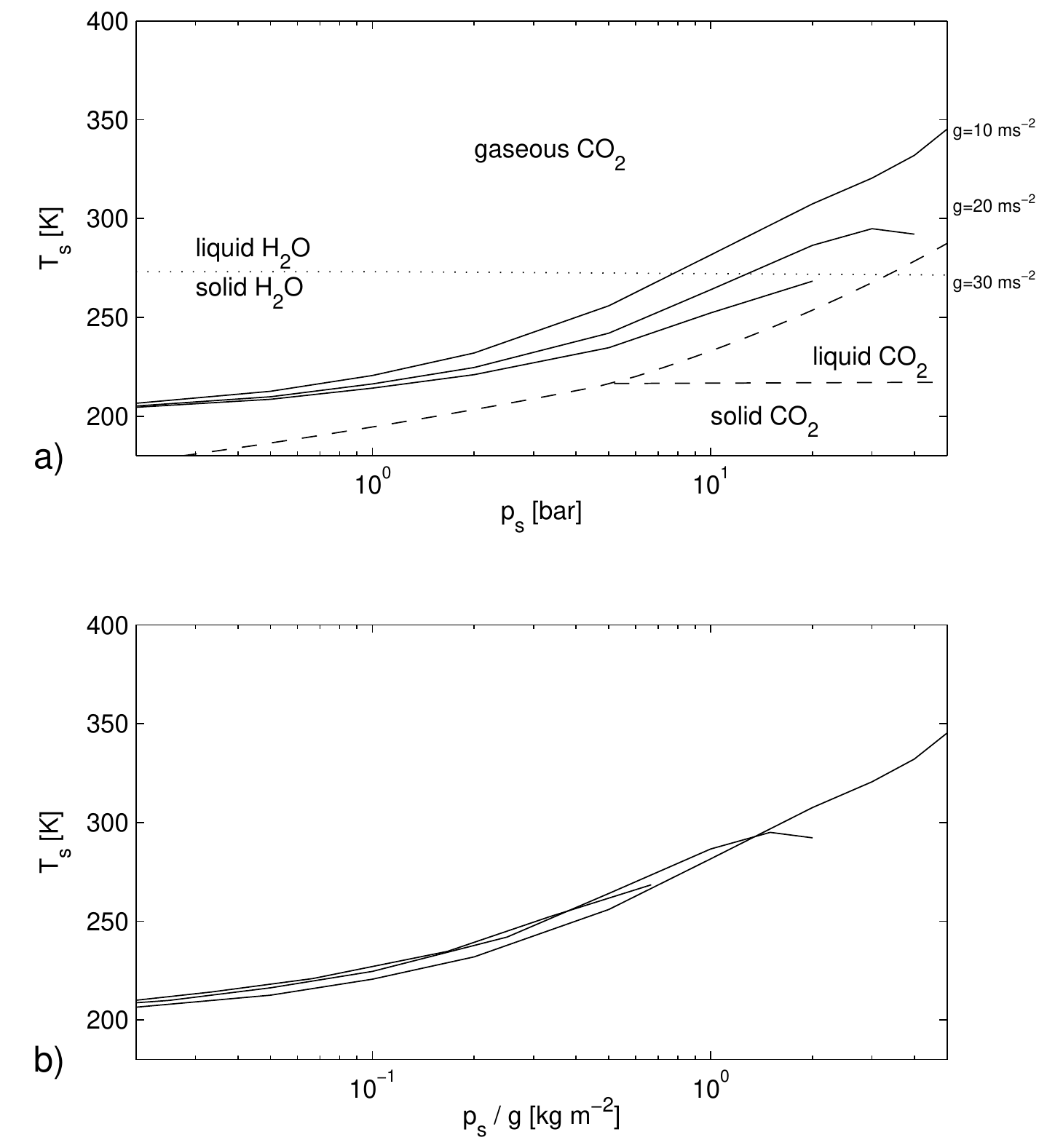}}
	\end{center}
	\caption{Surface temperature vs. surface pressure for clear pure CO$_2$ atmospheres under M-class insolation with varying surface gravity. Surface temperatures are plotted vs. a) pressure and b) CO$_2$ column density. Dashed and dotted lines in a) show CO$_2$ and H$_2$O phase transitions, respectively.}
	\label{fig:grav}
\end{figure}

\citet{Sotin2007} proposed a relationship between planetary mass and radius $(r\slash r_E)=\alpha (M\slash M_E)^\beta$, with $\alpha$ and $\beta$ equal to 1.0 and 0.274 for rocky planets and 1.262 and 0.275 for ocean planets, respectively. Given this relation and the current uncertainties in Gl581d's mass, we can expect the planet's surface gravity to be in the range $10 - 30$ m s$^{-2}$. In Figure \ref{fig:grav} we have plotted surface temperature vs. pressure for simulations with $g =$ 10, 20 and 30 m s$^{-2}$. For a given atmospheric pressure, the CO$_2$ column amount \mbox{$p_s \slash g$} (and therefore the total mass of the atmosphere) decreases with $g$. Hence the primary effect of increasing $g$, predictably, is to cool the surface. 

Variations in $g$ also change the adiabatic lapse rate, while leaving the CO$_2$ saturation pressure unaffected. Hence for a given CO$_2$ column amount the temperature profile varies with the gravity, which also influences the climate. These changes are important in determining the point at which maximum greenhouse warming occurs (see e.g. the $g = 30$ ms$^{-2}$ case in Figure \ref{fig:grav}a), but they do not strongly affect the surface temperature for a given CO$_2$ column amount at lower pressures. This can be seen in Figure \ref{fig:grav}b, where the same three curves are plotted vs. \mbox{$p_s \slash g$}. The atmosphere begins to collapse before the surface temperature reaches the water melting point only in the most conservative $g = $ 30 m s$^{-2}$ case.

\subsection{Surface albedo}\label{subsec:albedo}

\begin{figure}[h]
	\begin{center}
		{\includegraphics[width=3.5in]{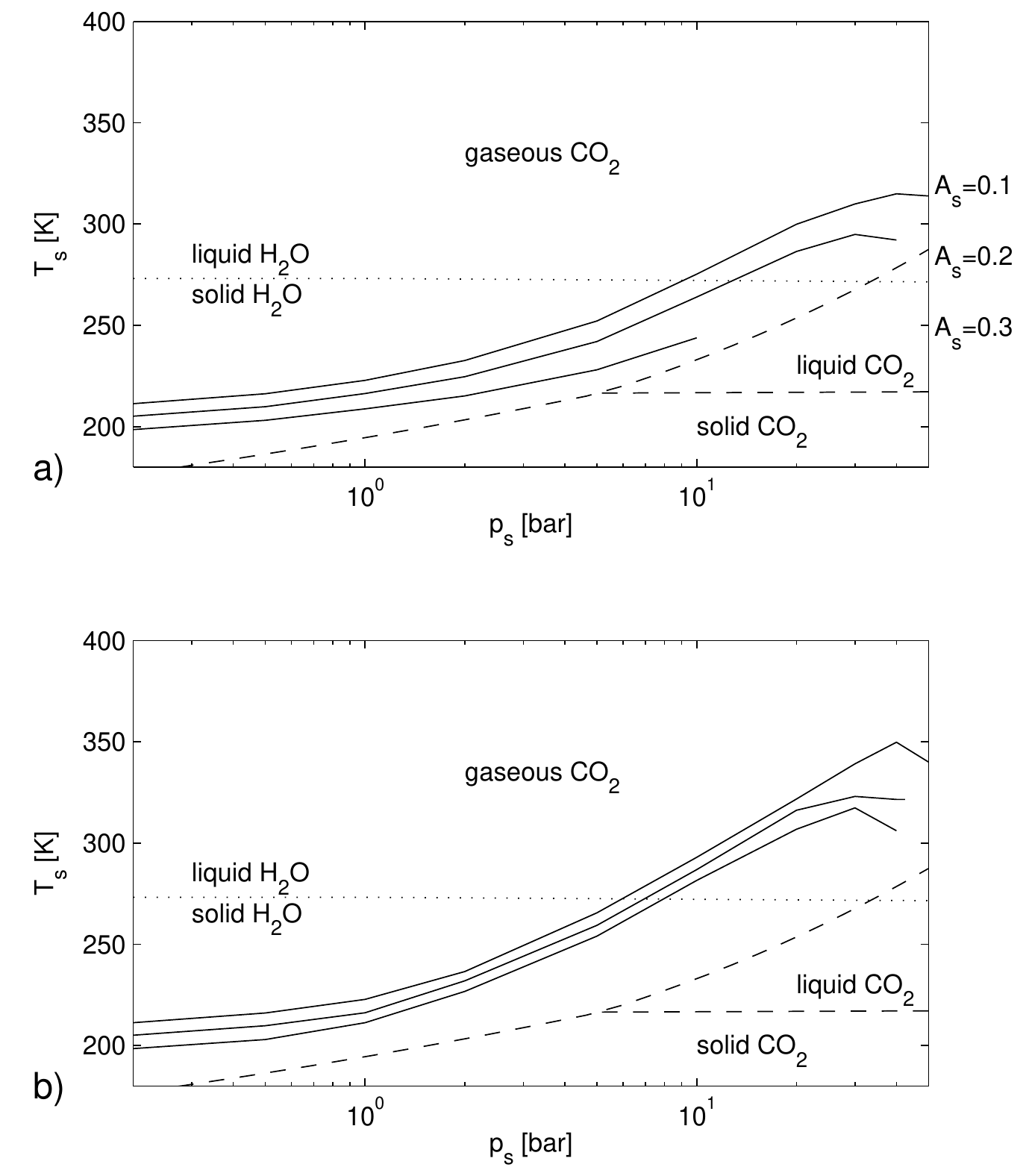}}
	\end{center}
	\caption{Surface temperature vs. surface pressure for  pure CO$_2$ atmospheres a) without clouds and b) with CO$_2$ clouds, $N_c=10^5$ kg$^{-1}$, under M-class insolation with varying surface albedo. Dashed and dotted lines show CO$_2$ and H$_2$O phase transitions, respectively.}
	\label{fig:albedo}
\end{figure}

Surface albedo has a dramatic effect on the surface temperatures of planets with thin atmospheres, but its importance might be expected to be less in dense or cloudy atmospheres. We tested the effect of variations in surface albedo in our calculations for $A_s =$ 0.1, 0.2 and 0.3. In Figures \ref{fig:albedo}a and b, we have plotted the surface temperature vs. pressure for clear-sky and $N_c=10^5$ kg$^{-1}$ CO$_2$ cloud cases, respectively. 

As can be seen, the difference in climate between $A_s=0.1$ and 0.3 is large in the clear-sky case. While the lower surface albedo leads to $T_s > 273$ K at 10 bar with a maximum \mbox{$T_s = 315$ K} at 40 bar, the higher surface albedo causes the atmosphere to collapse on the surface for $p_s > 10$ bar. However, in the cases with CO$_2$ cloud cover (Figure \ref{fig:albedo}b), the differences in surface temperature are generally smaller, and for all albedo values, surface temperatures above the water melting point are achieved before the surface pressure reaches 10 bar.  Hence if the atmosphere contains enough CO$_2$ to increase surface temperatures above 273 K and clouds of some kind are present, the effect of surface albedo on the climate should not be critical.

\subsection{N$_2$ partial pressure}\label{subsec:n2}

N$_2$ is an major constituent of the atmospheres of several rocky bodies in the Solar System. We tested the effects it would have on the climate of Gl581d by repeating the basic calculation of Section \ref{subsec:stellar} with added partial pressures of 1, 5 and 20 bars of N$_2$. As there are no important N$_2$ absorption bands in the visible or infrared, its primary effect on the climate is through a) additional Rayleigh scattering and b) changes in the specific heat capacity and mean molar mass of the atmosphere. We included both these effects in our calculations. N$_2$ also has a  warming effect in the Earth's atmosphere due to the broadening of CO$_2$ and H$_2$O spectral lines \citep{Goldblatt2009}. However, this effect is relatively small ($\sim$5 K for Earth conditions) and we are interested in conservative estimates of warming. Hence we neglected it here. 

\begin{figure}[h]
	\begin{center}
		{\includegraphics[width=3.5in]{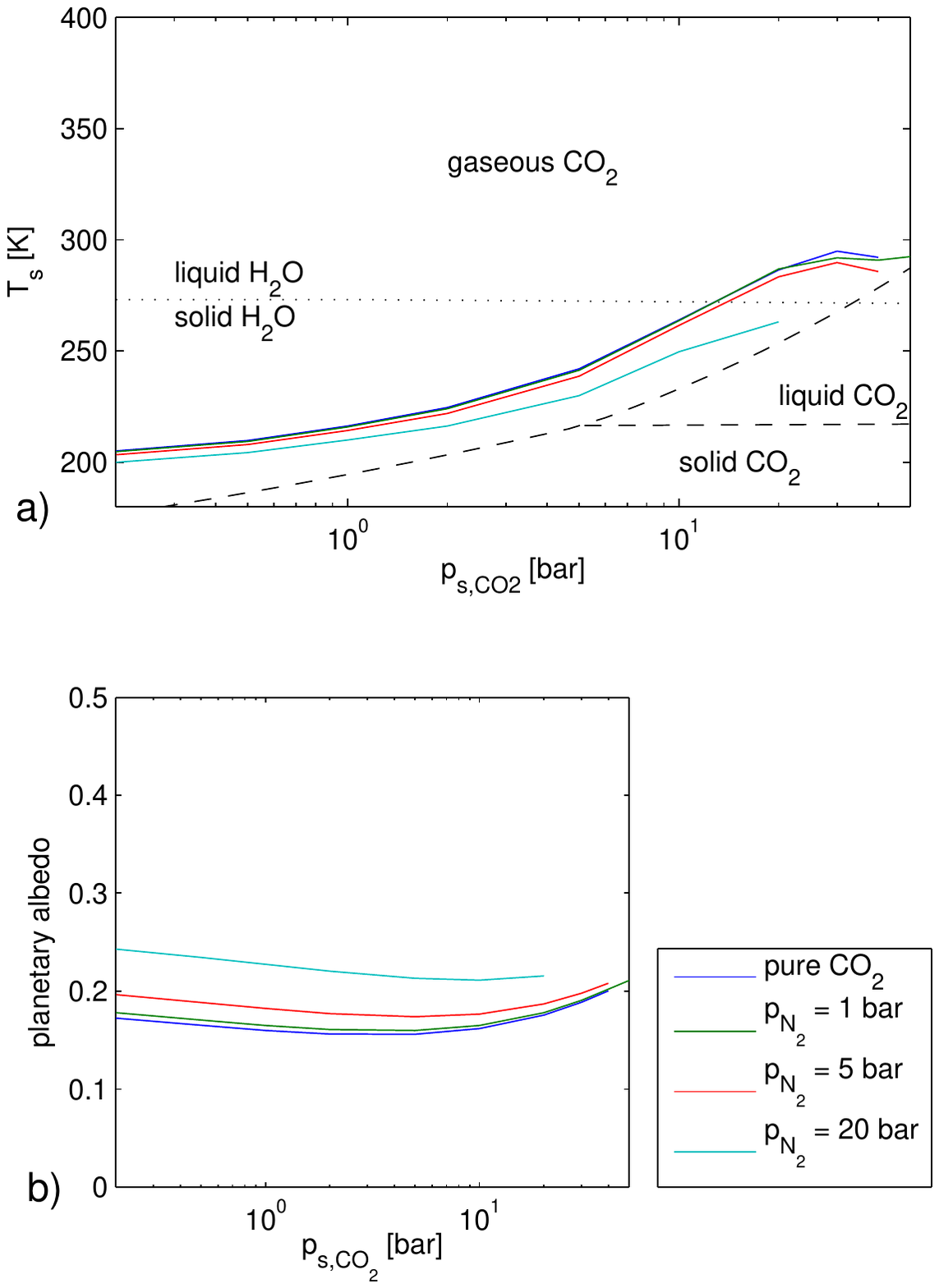}}
	\end{center}
	\caption{Surface temperature vs. surface pressure for clear mixed CO$_2$-N$_2$ atmospheres under M-class insolation.  Dashed and dotted lines show CO$_2$ and H$_2$O phase transitions, respectively. b) Planetary albedo vs. surface pressure for the same simulations as in a). In these figures, the $x$-axis shows the CO$_2$ partial pressure.}
	\label{fig:n2}
\end{figure}

As can be seen from Figure \ref{fig:n2}, the effect of N$_2$ on the surface temperature is relatively small. Only for $p_{N_2}= 20$ bar is the planetary albedo significantly increased above the value for pure CO$_2$ due to increased Rayleigh scattering. In this extreme case, the temperature fails to reach the water melting point, with the atmosphere collapsing after $p_s = 20$ bar. However, if N$_2$ warming were included, this difference would presumably be less significant.

\subsection{H$_2$O vapour and clouds}\label{subsec:h2ovap}

\begin{figure}[h]
	\begin{center}
		{\includegraphics[width=3.5in]{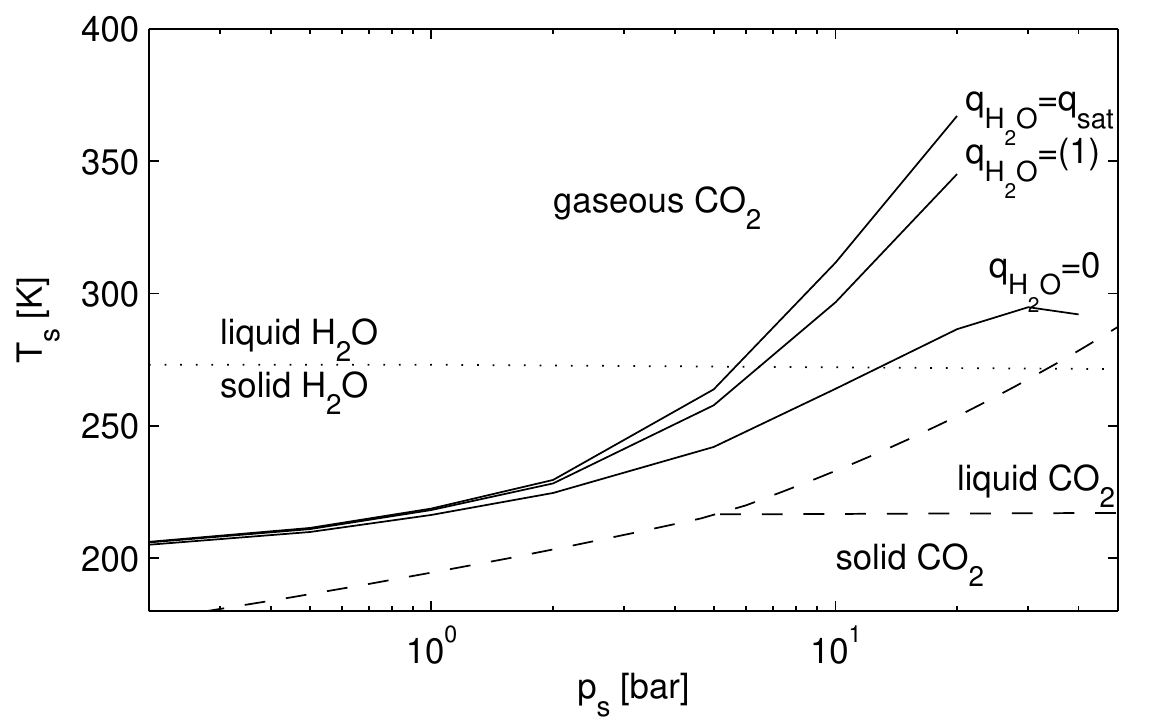}}
	\end{center}
	\caption{Surface temperature vs. surface pressure for clear mixed CO$_2$-H$_2$O atmospheres under M-class insolation.
 Dashed and dotted lines show CO$_2$ and H$_2$O phase transitions, respectively.}
	\label{fig:vap}
\end{figure}

Water plays an important and complex role in the radiative budget of Earth, through both direct vapour absorption and cloud absorption / scattering. Here we have studied its effects via the simple approach described in Section \ref{sec:method}. Figure \ref{fig:vap} shows runs performed with the relative humidity $RH=0$ at all levels, $RH=1$  at all levels, and  $RH$ defined as in (\ref{eq:RH}). As can be seen, increasing water vapour has the predictable effect of increasing the total warming, by an amount that increases with the surface temperature. This shifts the surface pressure at which we reach the water melting point down to 5 bar. Note that we neglect the effects of moist convection on the tropospheric lapse rate, so the CO$_2$+H$_2$O  results in Figure \ref{fig:vap} at the highest pressures likely overestimate the temperature somewhat.

\begin{figure}[h]
	\begin{center}
		{\includegraphics[width=3.5in]{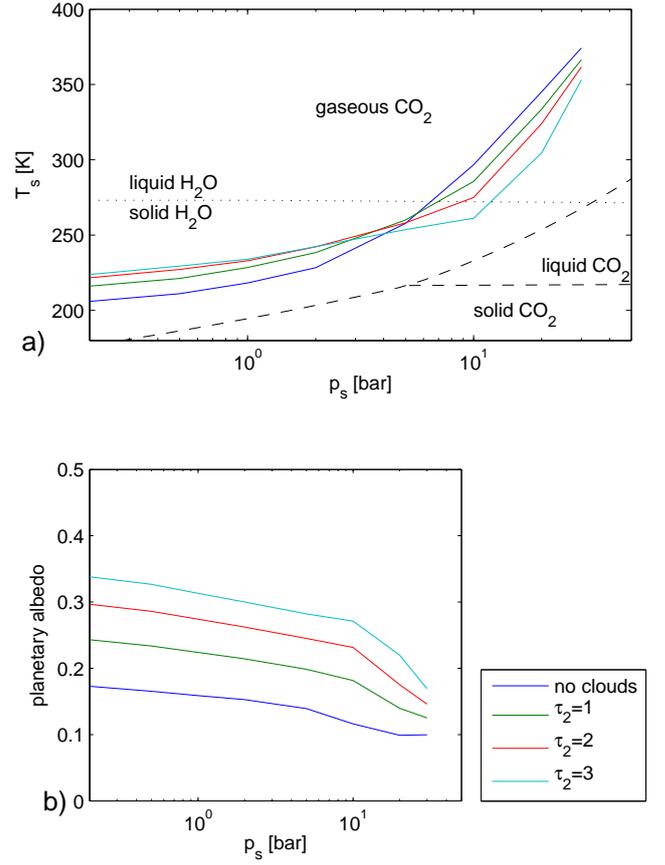}}
	\end{center}
	\caption{a) Surface temperature vs. surface pressure for mixed CO$_2$-H$_2$O atmospheres under M-class insolation with no CO$_2$ clouds and fixed H$_2$O cloud coverage. Dashed and dotted lines show CO$_2$ and H$_2$O phase transitions, respectively.  b) Planetary albedo vs. surface pressure for the same simulations as in a). In all cases optical depth is defined at the reference wavelength (1.5 $\mu$m).}
	\label{fig:clouds}
\end{figure}

Figure \ref{fig:clouds} shows simulations performed with fixed H$_2$O cloud layers in the troposphere  (see Section \ref{sec:method}) and varying vapor $RH$ humidity as defined in (\ref{eq:RH}). The planetary albedo increases with the cloud optical depth, but the amount by which it does so is lower at high pressures, because then atmospheric absorption/scattering above the clouds is more important. Counterintuitively, however, the clouds increase the surface temperature at low pressures and decrease it at high pressures. This effect can be explained by the fact that the atmosphere is relatively clear in some infrared spectral regions at low pressures. While an H$_2$O cloud layer reflects starlight, increasing the planetary albedo, it will also absorb outgoing infrared radiation from the surface, effectively acting as a `thermal blanket' and increasing the surface temperature. At high pressures, however, the lower atmosphere absorbs strongly in almost all infrared bands due to increased water vapour and the effects of CO$_2$ CIA. Hence the clouds have a minimal effect on the outgoing infrared radiation, but still increase the planetary albedo, which leads to a net cooling.

For higher values of  H$_2$O cloud optical depth, we found that at intermediate pressures, the atmosphere collapsed on the surface (see also Figure \ref{fig:conservative}). This suggests that sufficient H$_2$O cloud cover at low altitudes may limit the habitability of Gl581d. However, we used 100 \% fractional cloud coverage in these simulations, while on Earth, global low-level H$_2$O cloud coverage is generally less than 50 \%, with reported optical depths of 4.7 at 0.6 $\mu$m 
(\cite{kitzmann2010}, derived from results of \cite{Rossow1999} and \cite{Warren2007}). Hence our results for large optical depths may overestimate the cooling effect of the clouds at high pressures. For a more comprehensive study of the effects of H$_2$O clouds than we have given here, three-dimensional climate simulations including cloud microphysics  are necessary.

\subsection{Most conservative scenario}\label{subsec:worstcase}

\begin{figure}[h]
	\begin{center}
		{\includegraphics[width=3.5in]{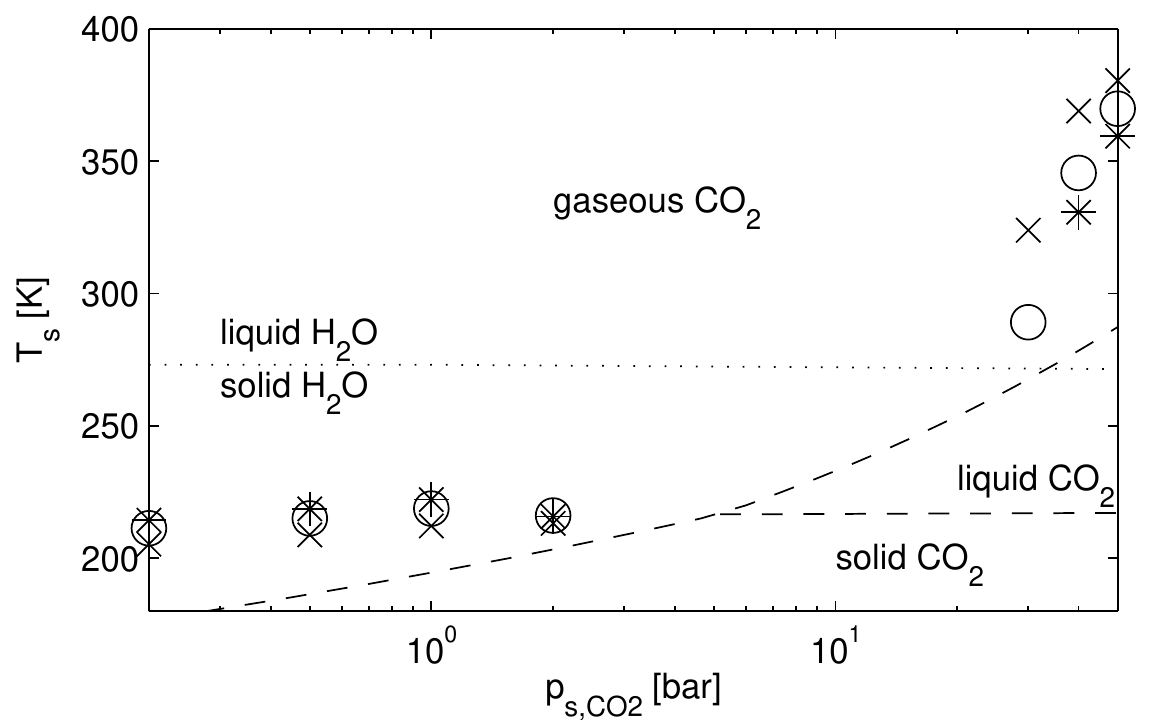}}
	\end{center}
	\caption{Surface temperature vs. surface pressure for the `most conservative case' described in the text. Dashed and dotted lines show CO$_2$ and H$_2$O phase transitions, respectively, while crosses, circles and stars indicate surface temperatures for atmospheres with H$_2$O clouds of opacity $\tau=$ 1, 2 and 3, respectively. In this figure, the $x$-axis shows the CO$_2$ partial pressure.}
	\label{fig:conservative}
\end{figure}

What happens if we combine the most conservative estimates for surface gravity, surface albedo, N$_2$ abundance and CO$_2$ cloud cover? Figure \ref{fig:conservative} shows a simulation that was performed with \mbox{$g = 30$ m s$^{-2}$}, $A_s = 0.3$, $p_{N_2}=20$ bar, minimal CO$_2$ cloud coverage ($N_c=10^3$ kg$^{-1}$) and water vapour present according to (\ref{eq:RH}). The H$_2$O cloud opacity was varied between $\tau$=1 and 3. 
As can be seen, in all three cases the atmosphere collapses at CO$_2$ pressures between 3 and 20 bar. 
Nonetheless, it can be seen that at higher pressures, the atmosphere is once again stable, with surface temperatures well above the water melting point at $p_{CO2}=$ 40 bar for these values of the H$_2$O cloud opacity. Given the initial assumption of CO$_2$ partial pressure of tens of bars, therefore, the outlook for the habitability of Gl581d would appear rather good.

\subsection{Three-dimensional equilibrium temperatures: airless planet case}\label{subsec:3Dnoatm}

While one-dimensional simulations of the type described here are useful because of their simplicity, they make the inevitable assumption of globally averaged stellar insolation. As discussed in \citet{selsis2007}, Gl581d is likely to be in a tidally resonant orbit, which means that there will be dramatic differences in insolation across the surface of the planet. To get an idea of the effect this could have on the climate, we performed some simple equilibrium temperature calculations in three dimensions.

\begin{figure}[h]
	\begin{center}
		{\includegraphics[width=3in]{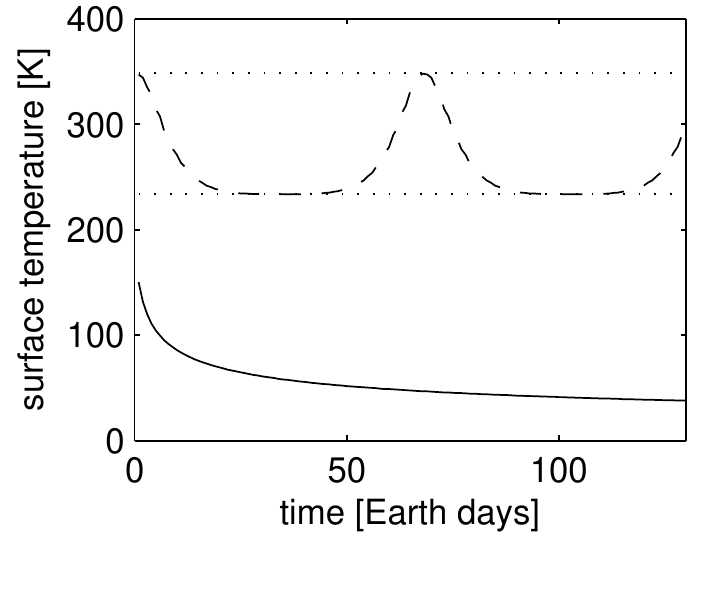}}
		{\includegraphics[width=3in]{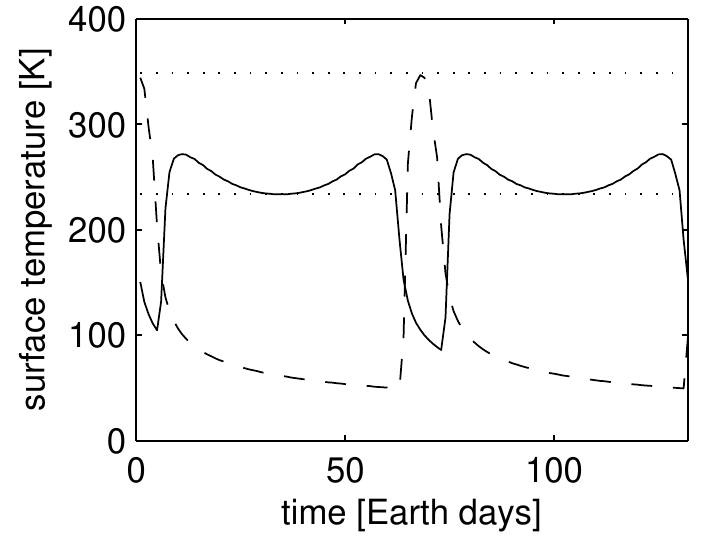}}
	\end{center}
	\caption{Snapshot surface temperature on the points of a hypothetical tidally resonant Gliese 581d
with minimal atmosphere and surface albedo A = 0.2 that are sub- (dashed line) and
anti-stellar (solid line) at periastron. Upper and lower dotted lines indicate the equilibrium sub-stellar temperature at periastron and apastron, respectively. The initial surface temperature in the simulations was 250 K at time = 0 days.}
	\label{fig:noatm}
\end{figure}

In Figure \ref{fig:noatm}, we have plotted the temporal evolution of surface temperature on the 0$^\circ$ and 180$^\circ$ longitude sides of an airless planet with the same orbit as Gl581d, zero obliquity, tidal resonances of 1:1 (left) and 2:1 (right), surface albedo $A_s=0.2$ and initial uniform surface temperature 250 K. Note that the `dark side' temperature in the plot on the left (solid line) is still decreasing toward 0 K in the simulation due to the finite thermal inertia that was given to the surface.

The dotted lines in the figure correspond to the equilibrium sub-stellar temperature at periastron
\begin{equation}
T_{pe}=\left(  \frac{S_0 (1-A_s)}{\sigma (1-e)^2 }  \right)^{1\slash 4}=344.0 \mbox{ K} 
\end{equation}
and apastron
\begin{equation}
T_{ap}=\left(  \frac{S_0 (1-A_s)}{\sigma (1+e)^2 }  \right)^{1\slash 4}=230.6 \mbox{ K}.
\end{equation}
As can be seen, at one location the local surface temperature can vary by as much as 300 K over the course of one orbit. This variability reduces for planets with atmospheres as a function of the surface pressure - on Venus, for example, where the diurnal period is 171 Earth days and the  pressure is 93 bars, the only significant surface temperature variation occurs due to topographic effects \citep{Mueller2008}. However, the decrease in variability of $T_s$ with pressure will be dependent in general on the specific atmospheric and orbital parameters of each planet. Hence further modeling is required before global temperature variations can be estimated for the hypothetical Gl581d atmospheres we have studied here. 

The second point of interest raised by this simple 3D model is that locally, the temperature on the planet rises well above 
the water melting point during the `warm season' (substellar point at periastron), despite the fact that there is no atmospheric warming. This raises an interesting question: could a mainly frozen planet in this orbit with a thin (p$_s > 6.1$ mbar) atmosphere contain a zone where liquid water forms seasonally? Two extreme cases can be envisaged, depending on the total amount of water present. If the fraction of water on the planet is small, the regions of low solar insolation are likely to act as cold traps in which water ice forms, leaving the zone in which liquid water is possible dry. If the planet's surface is covered by H$_2$O, however, there will be no limit to the amount of local evaporation that can occur. In this case, the local climate could potentially be quite interesting, although the high albedo of water ice would reduce the surface temperatures seen in Figure \ref{fig:noatm}. 

\section{Discussion}\label{sec:discuss}

Through one-dimensional radiative-convective climate modeling, we have found that as little as 5 bars of CO$_2$ ($\sim$2.5 bar equivalent column amount under Earth gravity) may be sufficient to maintain a global mean temperature above the melting point of water. For CO$_2$ partial pressures greater than 40 bar ($\sim$13 bar equivalent column amount under Earth gravity), we found that the mean surface temperature of Gliese 581d is high enough to allow surface liquid water even when the most conservative values of gravity, surface albedo and N$_2$ partial pressure are chosen, although  strong dependence on the total H$_2$O cloud opacity was observed.

We tested the dependence of surface temperature on a wide range of atmospheric and geophysical parameters. Because the M-class spectrum of Gliese 581 is red-shifted with respect to that of the Sun, heating due to near-IR absorption in the upper atmosphere increases in our simulations, while Rayleigh scattering is much less important. This is in agreement with the general habitability study of \citet{kasting1993}, but we find the difference to be even stronger than was reported there, due to the low effective temperature of Gl581. We also estimate an modest warming effect due to CO$_2$ clouds (up to 30 K at $p_{CO_2}$ = 50 bar), dependent on details of the cloud microphysics, although the efficiency of this process is reduced compared to the case for G-class stars. 

We find that increasing gravity primarily decreases the CO$_2$ column amount at which the maximum surface temperature occurs. Unsurprisingly, increased surface albedo decreases the mean surface temperature, although if there is cloud coverage, this effect is relatively small for the high CO$_2$ pressures at which the surface starts to become habitable. Increasing the partial pressure of N$_2$ has little effect below partial pressures $p_{N_2}$ of tens of bars. Finally, water vapour increases the surface temperature by a moderate amount below $T_s = 273 $ K, after which its role becomes increasingly important. We briefly assessed the impact of H$_2$O clouds, and found that they warmed (cooled) the surface at low (high) pressures, due to the interplay between their effects on incoming stellar and outgoing thermal radiation. 

As discussed by \citet{selsis2007}, N$_2$-CO$_2$-H$_2$O atmospheres likely only represent a fraction of the possible range for Earth-like / super-Earth extrasolar planets.  While the presence of radiatively inactive gases such as argon would be of little importance to the calculations discussed here, small amounts of other greenhouse gases such as methane would result in increased absorption, if such gases could remain chemically stable in the atmosphere. 
Unless the near-IR absorption of these gases were large enough to cause a strong temperature inversion in the upper atmosphere, this would further increase the surface temperature of the planet. The presence of other clouds or haze layers could alter the radiative balance in less predictable ways, depending on their scattering / absorption properties and height in the atmosphere. Unfortunately, effects of this kind are difficult to constrain further until direct atmospheric observations become available.

The results discussed here clearly have interesting implications, as they show that Gl581d could be the first discovered habitable exoplanet. They are also testable by future observations - dense CO$_2$ atmospheres have recognizable spectral signatures that could be detected, for example, using the proposed Darwin or TPF missions. However, there is an important limitation to this study: all simulations performed were one-dimensional. As we noted in Section \ref{subsec:3Dnoatm}, calculations of the \emph{globally averaged} surface temperature neglect variations due to local changes in stellar insolation (as well as topography and other effects). This has serious implications for dense, relatively cold CO$_2$ atmospheres, as if the dark side temperature of the planet is too low, CO$_2$ will condense on the surface. To test whether a dense atmosphere on Gl518d would be stable against the collapse of CO$_2$ in this way, we plan to repeat the calculations reported here in the future using a three-dimensional climate model.

\bibliography{allrefs}
\bibliographystyle{aa}

\begin{acknowledgements}
R. Wordsworth acknowledges funding from l'Institut de France, Fondation Cino et Simone Del Duca.
F. Selsis  acknowledges support from the European Research Council (Starting Grant 209622: E$_3$ARTHs).
\end{acknowledgements}

\section{Appendix}

To assess the importance of the non-ideal behaviour of CO$_2$ gas at high pressures in our calculations, 
we used the modified Redlich-Kwong equation of state proposed by \citet{Li2006}. The Redlich-Kwong equation takes the form
\begin{equation}
p=\frac{RT}{v-b} - \frac{a}{v(v+b)\sqrt{T}}\label{eq:RK},
\end{equation}
where $p$, $T$, $v$ and $R$ are pressure, temperature, specific volume and the ideal gas constant, respectively. 
Various definitions of $a$ and $b$ are possible; in \citet{Li2006} $a=a_1(T)+a_2(T)p$, with $a_1$ and $a_2$ polynomial functions of temperature, while $b=0.08664R T_c \slash p_c$, with $T_c$ and $p_c$ the critical point temperature and pressure for CO$_2$. Equation (\ref{eq:RK}) can be written as a cubic equation in terms of $v$ and hence is solvable analytically. The inverse compressibility factor $\beta=\mathcal Z^{-1}=RT \slash pv$ calculated from (\ref{eq:RK}) agrees approximately with the values stated in \citet{Kasting1991} for the triple point and saturated CO$_2$ at 0$^\circ$C. This is unsurprising, as the equation used there was of the same basic form. 

To get a first-order estimate of the effect of non-ideality on the tropospheric lapse rate (and hence on the climate), we approximated (\ref{eq:RK}) to a virial form as
\begin{equation}
pv \slash RT\approx 1+B/v \label{eq:RKvirial}
\end{equation}
with 
\begin{equation}
B(T)=b - \frac{a}{RT^{3\slash 2}}.
\end{equation}

Given the definition of specific heat capacity in \citet{Kasting1991}
\begin{equation}
c_p(p,T) = c_p(0,T)-T\int^p_0 \frac{\partial^2 v}{\partial T^2}\vline_p dp' 
\end{equation}
it is clear that in the virial approximation, to leading order
\begin{equation}
c_p(p,T) =  c_p(0,T)-TpB''(T) 
\end{equation}

Various formulations for $c_p(0,T)$ were considered, including the quadratic function given in  \citet{Kasting1991} 
and the power law function of \citet{Leb2010}. 
We found that the latter expression, which is defined as $c_p(0,T) = 1000\left(T \slash 460 \right)^{0.35}$, had the greatest accuracy vs. reference data \citep{CRC2000}.  Using this and calculating $B''(T)$ numerically by finite differences, we found for  $T = 250$ K  and $p = $ 0, 50 bar that $c_p = 795.9$ and $825.0$ J K$^{-1}$ kg$^{-1}$, respectively. 
Climate calculations performed with fixed specific heat capacities of 750 and 850 J K$^{-1}$ kg$^{-1}$ showed maximum temperature differences of only 2.2 K over the entire range 0.2 - 50 bar, with the latter value of $c_p$ giving colder results. Hence we decided to ignore variations in $c_p$ with altitude in our calculations, and used the upper limit $c_p = $ 850 J K$^{-1}$ kg$^{-1}$ everywhere in the pure CO$_2$ calculations. For mixed CO$_2$-N$_2$ atmospheres, a weighted mean value for $c_p$ was used, with the heat capacity of N$_2$ taken to be a constant 1003 J K$^{-1}$ kg$^{-1}$.

\end{document}